\documentstyle[12pt]{article}
\newcommand{\n}{\nonumber}
\def\p{i+1/2}
\def\m{i-1/2}
\begin{document} 
\begin{center}
 \begin{bf}
  HYDRODYNAMICAL EVOLUTION\\ OF A QUARK-GLUON PLASMA DROP WITH BOUNDARY
 \end{bf}
~\\~\\
     S.P. BARANOV, L.V. FIL'KOV and N.A. LOKTIONOVA\\
{\sl P.N.Lebedev Physical Institute, Moscow 117924, Russia}\\~\\~\\
\parbox{14cm}{\begin{small}
                 \quad We consider the evolution of a limited Quark-Gluon
 Plasma (QGP)drop in the framework of relativistic hydrodynamics. 
In the presence of the
 boundary, the expanding and cooling of QGP drop 
may appear as oscillating processes,
 which realize via multiple traveling of rarefaction and compression waves
through the plasma and leads to increasing of plasma life-time.
 Within this picture, deflagration seems to be the main possible way of
 hadronization. The slow cooling mechanism also leads to a significant
 enhancement of photon yield in comparison with Bjorken scenario.
\end{small}}\\~\\~\\
\end{center}
\begin{large}
\begin{center}
\begin{bf}
 1. Introduction\\
\end{bf}
\end{center}

 When discussing the signals of a quark-gluon plasma produced in heavy ion
 collisions, particularly the ones connected with thermal photons or lepton
 pairs \cite{phot2}, it is necessary to have an adequate underlying picture
 of the space-time evolution of plasma. Moreover, the very characteristic
 details of the evolution of plasma may be considered as signals by their own
 \cite{F}.

 In principle, it is generally accepted that relativistic hydrodynamics is able
 to provide the wanted description, but the exact solution of the mathematical
 problem is rather difficult. The approach, which is most widely exploited
 in this context, if not to say the only one, is known as Bjorken model
 \cite{Bj}. This model is based on the assumption of a space-time scaling,
 which means that all processes in a hot strong-interacting system do only
 depend on the local proper time $\tau$. The assumption of space-time scaling
 leads to major simplifications and allows to obtain an analytical solution of
 the hydrodynamical equations. At a further step, this scaling solution is
 often used as a ground for more sophisticated models \cite{Ru1,Ru2,Bj1}
 where the transverse motion of matter is taken into account as a correction
 to its scaling-like longitudinal expansion.

 However, this model is only an idealization, which ignores some important
 features of the real world. For example, it is the fact that the QGP drop is,
 anyway, limited, and so there must exist a boundary, which separates two
 phases with different equations of state.

 Our own study is based upon a direct numerical analysis of hydrodynamical
 equations together with the boundary conditions between plasma and vacuum.
 In the present work we have considered the space-time evolution of
 a spherically-symmetric QGP drop and have tested various sets of initial
 conditions.

 A similar approach, also applied to a spherically symmetric QGP drop, has been
 proposed in ref. \cite{Pratt}. The author of \cite{Pratt} mainly focuses his
 attention on the effects of pion-pion Bose-Einstein correlations and compares
 his predictions with Bjorken Model. This is well complementary to our present
 consideration, where the emission of thermal photons is revised. In addition,
 we include in our analysis various ways of hadronization, both equilibrium
 and nonequilibrium, whereas the paper \cite{Pratt} deals with only one type
 of phase transitions, the deflagration.

 Having the {\it a priori} space-time scaling given up, we get a new,
 qualitatively different view of the evolution of plasma.
 Among the noticeable consequences of our analysis is the fact that
 supersonic phase transitions from QGP to hadrons are impossible, and this
 may strongly affect the eventual proportion between different particles,
 which appear in the course of hadronization \cite{BF1,BF}.
 Another important observation is the enhanced yield of thermal photons,
 which exceeds the traditional estimations by about one order of magnitude.
 Our results remain valid in a wide range of model parameters.

 The outline of the paper is the following. In sect. 2 we present the basic
 equations of hydrodynamics and the equation of state of the deconfined matter.
 In sect. 3 we briefly discuss the numerical scheme to integrate the dynamical
 equations with boundary conditions. In sect. 4 we analyze the evolution of
 a QGP drop as it appears in our approach and point out some essential features
 of the dynamics. In sect. 5 we calculate the production of thermal photons.
 Our main findings and conclusions are summarized in sect. 6.\\~\\
\begin{center}
\begin{bf}
 2. Equations of motion and equations of state\\
\end{bf}
\end{center}

 To describe the evolution of matter created in a relativistic collision of
 heavy nuclei one commonly uses the hydrodynamical approach \cite{LL,ref2}.
 The equations of motion represent the momentum and energy conservation. In the
 absence of dissipative processes (diffusion, viscosity, heat conductivity)
 they read
\begin{equation}
{\partial}_{\mu}T^{\mu\nu}=0, \label{1}
\end{equation}
 where the energy-momentum tensor
\begin{equation}
T^{\mu\nu}=(\varepsilon +p)\,u^{\mu}u^{\nu}-pg^{\mu\nu}. \label{2}
\end{equation}
 Here $\varepsilon$ is the energy density, $p$ is the pressure,
 $u^{\mu}$ is the 4-velocity, and $g^{\mu\nu}$ is the metric tensor.

 For the spherically symmetric case considered here it is convenient to rewrite
 the above equations (1),(2) in an explicitly symmetric form:
\begin {eqnarray}
\partial_t{\bf u} +{1\over{r^2}}\,\partial_r (r^2 {\bf f(u)}) = 0,\,
{\bf u}\equiv  {w\gamma^2 -p\choose{w\gamma^2 v_r}},\,
{\bf f(u)}\equiv {(w\gamma^2 -p)v_r + pv_r\choose{w\gamma^2 v^2_r +p}},
\end {eqnarray}
 where $w=\varepsilon+p$ is the enthalpy density, $v_r$ is the radial component
 of the velocity, and $\gamma=(1-v_r^2)^{-1/2}$.

 The dynamics of a QGP drop is now completely defined by the initial conditions,
 the boundary conditions and the equation of state, which relates $p$ and
 $\varepsilon$.

 We adopt the MIT-motivated equation of state of plasma \cite{mit} and get for
 massless quarks and gluons
\begin{eqnarray}
&& p =\alpha T^4 -B, \n \\[-2mm]
&& \label{3}            \\[-2mm]
&& \varepsilon = 3\alpha T^4+B, \n
\end{eqnarray}
 where $T$ is the temperature and $B$ is a phenomenological constant
 corresponding to the energy density and the pressure of the QCD vacuum. The
 parameter $\alpha$ defines the number of degrees of freedom in QGP and is
 taken  $ \alpha = \pi^2/90 (2\times 8 + 2\times 2\times 3\times N_f\times 
 \frac{7}{8}) \approx 23\pi^2/45,$  where $N_f = 2.5$ for taking into account
 u- , d-, s-quarks.

 Since the QGP drop is assumed to be of limited size it is surrounded by the
 hadronic vacuum with the equation of state
\begin{equation}
\varepsilon_{vac}=p_{vac}=0.\label{vac}
\end{equation}
 We will recall this equation when formulating the boundary conditions.\\~\\
\begin{center}
\begin{bf}
 3. Numerically integrating the hydrodynamical equations\\
\end{bf}
\end{center}

 Numerical algorithms aimed at the solution of the hydrodynamical problems are
 being developed for many years. Among the various approaches described in the
 literature we prefer the nonlinear monotonic transport scheme FCT
 (Flux-Corrected Transport) \cite{FCT1,FCT2,FCT3} developed by J.P.Boris and
 D.L.Book in the framework of Eulerian grid methods. It has been shown to apply
 to a wide class of physical systems and is found quite accurate and reliable.

 Let us remind the guideline of this method.

 Consider one-dimensional continuity equation for the mass density flux:
\begin{equation}
\partial _t \rho + \partial _x (\rho v) = 0.
\end{equation}
 Let $\rho_i^0$ denote the density in the $i^{th}$ point of the grid at a given
 time. An explicit scheme to calculate the density $\tilde\rho_i$ at the next
 time step may be presented as follows:
\begin{eqnarray}
\tilde\rho_i &=&\rho^0_i -\epsilon_{\p}(\rho^0_{i+1}+\rho^0_i)
       +\epsilon_{\m}(\rho^0_{i-1}+\rho^0_i) \label{rho} + \nonumber \\
   & & \nu_{\p}(\rho^0_{i+1}-\rho^0_i)+\nu_{\m}(\rho^0_{i-1}-\rho^0_i).
\end{eqnarray}
 This form observes the mass conservation.
 Here the quantities with half-integer indices $i\pm 1/2$ correspond to
 arithmetic-mean values of the ones taken in the points $i$ and $i\pm 1$, and
 $\epsilon_{i\pm 1/2}\simeq (1/2)\,v_{i\pm 1/2}\,(\delta t/\delta x)$ with
 $\delta t$ being the time step and $\delta x$ the space constant of the grid.
 The terms with $\nu$ stand for an artificial diffusion and are introduced to
 guarantee the positivity of mass density, which demands
\begin{equation}
\nu_{i\pm 1} \geq |\epsilon_{i\pm 1}|\qquad \mbox{for all}\,\,i
\end{equation}
 To compensate this overstrong diffusion one applies an antidiffusion
correction:
\begin{equation}
\rho^1_i = \tilde\rho_i -\mu_{\p}(\tilde\rho_{i+1} -\tilde\rho_i)
                        -\mu_{\m}(\tilde\rho_{i-1} -\tilde\rho_i),
\end{equation}
 with $\mu_{i\pm 1/2}$ being positive coefficients.
 This step removes extra smoothing but no longer guarantees the positivity
 of the newly computed densities $\rho^1_i$. To restore the positivity and
 the stability of the eventual solution the antidiffusion fluxes
 $f_{i\pm 1/2} \equiv \mu_{i\pm 1/2}(\tilde\rho_{i\pm 1} -\tilde\rho_i)$
 need to be corrected. In the simplest form one can take
$\mu_{i\pm 1/2} = \nu_{i\pm 1/2} - |\epsilon_{i\pm 1/2}|.$
 Eliminating the residual diffusion $(\nu-\mu)$ is only possible with nonlinear
 corrections adjusted to the local behaviour of the solution.

 The fundamental requirement of the FCT method is that the antidiffusion step
 must neither generate new extrema in the solution nor enhance the ones already
 existing. This is observed with the following prescription:
\begin{eqnarray}
\rho^1_i &=& \tilde\rho_i - f^{cor}_{\p} - f^{cor}_{\m}, \n \\
 f^{cor}_{i\pm 1/2} &=& S\,\mbox{max}\{0,\,\,
  \mbox{min} [S(\tilde\rho_{i\pm 2} -\tilde\rho_{i\pm 1}),\,\,
 |f_{i\pm 1/2}|,\,\,S(\tilde\rho_i -\tilde\rho_{i\mp 1})]\}, \label{limit}\\
 |S| &=& 1,\qquad
 \mbox{sign}\,S = \mbox{sign}(\tilde\rho_{i\pm 1} -\tilde\rho_i). \n
\end{eqnarray}
 The condition (\ref{limit}) is constructed in such a way that it takes
 into account all possible combinations of the signs of local gradients
 of the physical solution. A detailed description of various transport,
 antidiffusion and correction schemes may be found in \cite{FCT3}.
 Like it is done in the original papers \cite{FCT2,FCT3}, we also use the
 time-splitting procedure to calculate the velocities.

 An essential point in our analysis is the presence of the moving phase
 boundary, which greatly complicates the consideration. In general, Lagrangian
 grids are better suited for this kind of problems. Since the grid cells are
 made moving together with the matter, one can identify the intercell boundary
 with the phase boundary, thus avoiding the ambiguous situation of mixing two
 different equations of state in one cell. To fit in with the FCT method
 we have adopted a combined scheme.

 All the internal cells of our grid are of Eulerian type and have definite
 unchangeable size (it will be refered to as "standard size"). On the contrary,
 the cell adjacent to the phase boundary is of Lagrangian type, with its outer
 face moving together with the boundary. Under the evolution of the QGP drop
 the size of this cell changes continuously. If, due to the expansion of the
 drop, the length of the surface cell becomes larger than 1.5 standard lengths,
 the cell splits into two new cells. The one, which does not touch the phase
 boundary, gets the full standard size. It now belongs to the Eulerian part
 of the grid. The other cell, which is adjacent to the phase boundary, takes
 the rest of the total length and continues its life as Lagrangian cell. In the
 opposite case, if the length of the surface cell reduces to a quantity less
 than 0.5 standard lengths, it fuses with the neighboring Eulerian cell, thus
 forming a Lagrangian cell of extended size.

 All the internal Eulerian cells are treated according to the standard FCT
 algorithm. The parameters of the boundary Lagrangian cell are found from
 the momentum and the energy conservation. The incoming and outcoming fluxes
 of the momentum and the energy for the Lagrangian cell are known from the
 parameters of the neighboring Eulerian cells on the inner side and from the
 equation of state of vacuum (\ref{vac}) on the outer side. Given these fluxes,
we
 calculate the new values of the momentum and the energy densities. Then
 we extract from these quantities the velocity of plasma. This velocity
 is assumed to be the velocity of the boundary of the QGP drop. Given the
 velocity, we calculate the new size of the Lagrangian cell and recalculate
 the momentum and the energy densities according to the changed size.

 The calculated values of the densities and the volume of the Lagrangian cell
 may then be corrected in such a way that to achieve an exact conservation of
 the energy. This means
\begin{equation}
\varepsilon_N V_N =
\sum_{i=1}^{N}\varepsilon_i V_i -\sum_{i=1}^{N-1}\varepsilon_i V_i
\equiv E_{(t=0)} -\sum_{i=1}^{N-1}\varepsilon_i V_i ,
\end{equation}
 where $E_{(t=0)}$ is the initial total energy of the system. This equality
 may be used either to correct the volume $V_N$ of the Lagrangian cell
 (if to think the energy density $\varepsilon_N$ is known correctly) or
 to correct the energy density (if to think the volume is known)
 or to make a weighted correction for both these quantities. The latter
 case is realized in our paper.
 This scheme was tested in 1-dimensional case with very simple
 Lagrangian method \cite{dis}.\\~\\
\begin{center}
\begin{bf}
 4. The dynamics of a QGP drop\\
\end{bf}
\end{center}

 In the present section we consider the dynamical properties of the evolution
 of a QGP drop. The initial conditions for this problem must be derived from
 the dynamics of nucleus-nucleus collisions. In fact it leaves a large piece of
 freedom. We do not pretend to give precise and detailed predictions on the
 behaviour of plasma in real experimental conditions. Our purpose is rather to
 give a qualitative description, which reveals the characteristic features of
 a new mechanism of expansion and cooling.

 For illustrative purposes, we consider a spherically symmetric baryonless
 drop of quark-gluon plasma. The main conclusions derived from our analysis
 would neither depend on the exact shape of a QGP drop nor on the details of
 the equation of state such as nonzero baryon density. Instead, it is the
 presence of the boundary, which is the origin and the reason of the
 distinctive properties of our model.

 So we consider a spherical drop at a constant temperature $T_0$ and pressure
 $p_0$ in the initial moment. The initial velocity is assumed to rise linearly
 from $|{\bf v}|=0$ at the center of the drop to $|{\bf v}|=v_0$ at the
 boundary. We have tried widely different values of $p_0$ and $v_0$. We have
 also tested the sensitivity of the results on the choice of the bag constant
 $B$ in the equations of state and to the strength of the surface tension,
 which determines the mechanical stability of the drop against breaking.

 Consider first the case of zero initial velocity $v_0=0$ (fig. 1)
and following parameters: radius $r_0 = 3~fm$, $p_0 = 6.5~GeV/fm^3$, 
 $B=0.2~GeV/fm^3$.
 The expansion of a QGP drop starts from the nearby-surface regions.
 The overfall of pressure (from plasma to vacuum) leads to the formation
 of a rarefaction wave, which propagates from the surface of the drop to
 its center (fig. 1 a,b,c). The velocity of this wave is the speed of sound.
 This wave puts the plasma in motion, thus causing the process of expansion.
The continuous expansion of the QGP drop during all this
 time results in a monotonic decrease of the pressure and the temperature in
 the whole of the drop (fig. 1 a-h).

 If the initial velocity is high ($v_0\ge v_s$,
 $v_s$ being the speed of sound) the evolution starts from the
  creation of  excess pressure region
 at the boundary of the QGP drop. This leads to a formation of a
 compression wave, which also moves from the surface of the drop
 to its center with the speed of sound. In the same time the QGP drop expands as
 a whole, so that the pressure and the temperature decrease as in the previous
 case. When the compression wave reaches the center of the drop and reflects
 from it, the evolution of the drop becomes very similar to the case of zero
 initial velocity.

 The expansion of a QGP drop may be induced either by the rarefaction
 wave (if $v_0=0$) or the initial motion ($v_0 > 0$), or both these reasons,
 but the behaviour of plasma looks the same in the sense that the expansion and
 cooling are always accompanied by the propagation of waves.

 Depending on the initial amount of energy (or the initial temperature) the
 expansion may take different time. The compression waves may pass through the
 drop several times, reflecting successively from the center and from the
 boundary. The expansion and cooling of plasma carry on until the pressure
 falls below zero or the conditions for a phase transition are reached.
 What happens next depends on the mechanical and thermodynamical stability
 of the plasma. For the time, let us forget about phase transitions and
 concentrate on a purely hydrodynamical problem.

 The mechanical stability of plasma depends on the strength of its surface
 tension. The estimations of this parameter are uncertain, and hence we will
 consider two opposite extreme cases.

 First, assume that the surface tension is always strong enough to prevent
 the QGP drop from breaking into smaller droplets. Then the expansion of plasma
 by inertia may lead to states with negative pressure. If the initial pressure
 was small compared to the bag constant $(p_0<B/2)$, the negative
 pressure appears for the first time at the moment when the rarefaction wave
 reaches the center of the QGP drop. Since that moment the region of negative
 pressure spreads with the reflected wave towards the drop boundary. The moment
 when it reaches the surface of the drop corresponds to the maximal expansion
 of plasma and to its deepest overcooling. At this time the pressure is almost
 uniformly negative in all the volume of the drop and its value is approximately
 equal to the initial pressure taken with the opposite sign: $p_{0}\simeq -p_0$.
 After that the drop starts to squeeze. At the end of the period of squeezing
 plasma returns to the state, which is similar to the initial one, both in its
 pressure (or temperature) and the occupied volume.

 If the initial pressure is high $(p_0>B/2)$, the process of expansion and
 cooling takes a longer time.
As result of reflaction of a plasma flow from
 the drop boundary, some excess of pressure is formed near the boundary. It
 creates the compression wave passing through the plasma
 drop many times reflecting successively from its center and from its surface
 at the background of the more strong rarefaction process.
 This compression wave slows down the expansion of plasma and eventually stops
 it. It is remarkable that even at the moment of maximal overcooling, the
 negative pressure can never fall below a definite value, which depends on the
 bag constant $B$ but does not depend on the initial pressure $p_0$:
 \,\,\, $p>p_{min},\,\,\,p_{min}\simeq -B/2$ (fig. 1 j).
 The subsequent squeezing of plasma restores positive pressure in the whole of
 the drop. However, the value of this positive pressure $p'_0$ is not equal to
 the initial value $p_0$, but is nearly the minimal (negative) pressure taken
 with the opposite sign: $p'_0\simeq -p_{min}$ (fig. 1 l). Neither does the
 occupied volume contract to the initial size.

 If to allow a further evolution of plasma, one would see a second period of
 expanding and squeezing. It essentially resembles the first period, but
 corresponds to a reduced initial pressure, i.e. $p'_0$ instead of $p_0$
 (figs. 1 l). At the end of this period plasma comes to a state with
 $p''_0\simeq  p'_0$, as is typical for processes starting from relatively low
 pressure. All the subsequent periods are simply a repetition of this picture.

 On the other hand, if the plasma surface tension is weak the evolution of the
 drop comes to end as soon as the pressure falls to zero. Then the creation of
 vacuum bubbles (or breaks) comes into play. The breaks prevent the temperature
 and the pressure from further decreasing and "freeze" the state with
 $p\simeq 0$.

 Most probably, the vacuum bubbles will break the QGP drop into smaller
 droplets, but their summary volume must conserve. The volume cannot increase
 because this would lead to further decreasing pressure and will simply end up
 with new breaks. The volume cannot decrease because there is no force, which
 could cause squeezing of plasma, while the motion by inertia tends to move
 the drops away from each other. Hence, the primordial QGP drop will transform
 into a sponge-like object or into a cloud of smaller droplets. The pressure
 in this system freezes at the $p\simeq 0$ level.

 It is important to point out that the dynamical limit for overcooling
 $p_{min}\simeq -B/2$ remains valid in all the cases.
 The existence of this limit restricts the possible ways of hadronization.
 An explanatory overview of all types of phase transitions may be found in the
 textbook \cite{LL}, and so we will not repeat it here.
 According to the estimations made in \cite{BF,BF1} the supersonic
 condensation and the detonation may only occur if $p < -B/2$.
 Since this degree of the overcooling can never be achieved, no matter
 what kind of initial conditions are chosen, we are left with slow processes
 like equilibrium phase transition and deflagration. The deflagration must
 be considered the most realistic scenario for the hadronization of plasma.

 Now let us take the deflagration into account. By the definition \cite{LL},
 the deflagration is a slow process, which spreads with a velocity much less
 than the speed of sound. There are no reasons to accept that the plasma can
 convert into hadrons in a moment, once the pressure has fallen down to
 the equilibrium threshold. Moreover,
 the deflagration
 takes place only at the  surface of plasma drop, but not in its inner
 regions.

 In the present consideration we have simulated the deflagration as a process,
 which starts at the outer surface of the drop and spreads towards the 
center
 with a constant velocity $v_{def}=0.1~c$ \cite{BF1}.
 The deflagration proceeds during the time when the necessary condition
 $0<p<p_{th}$ is satisfied, where the threshold value $p_{th}$ has been
 estimated in \cite{BF,BF1}: $p_{th}=30~MeV/fm^3$.

 Fig. 2 illustrates the evolution of a QGP drop with the account of
 deflagration and $p_0 = 2.56 GeV/fm^{-3}$ 
(assuming the negative pressure is allowed). Hadronization
 reduces the mass of plasma, but proceeds with a relatively small velocity.
 One can see several oscillations of the drop before the termination of the
 phase transformation. If the initial drop would break into smaller droplets
 (assuming the negative pressure is unstable) the deflagration would proceed
 faster because of enlarged total surface of the system. However, it takes
 a rather long time to reach even the state, where the 
 deflagration is possible. 
This fact
 has a direct consequence in the enhanced yield of thermal photons and lepton
 pairs.\\~\\
\begin{center}
\begin{bf}
 5. The yield of thermal photons\\
\end{bf}
\end{center}

 To estimate the yield of photons from QGP we recall the analytical results of
 ref. \cite{neubert}. The emission of photons is mainly due to two elementary
 processes, i.e. the gluon Compton scattering and the annihilation of quarks:
\begin {eqnarray}
q(p_1)+g(p_2) \to \gamma(k)+q(k'), \\
q(p_1)+\overline q(p_2) \to \gamma(k)+g(k').
\end{eqnarray}
 Here the letters in the parentheses denote the momenta of the particles.
 In the lowest QCD order, the spin- and color-averaged matrix elements for
 these processes read:
\begin {eqnarray}
\vert T_{qg}\vert^2 &=& -{16\pi^2\over 3} Q_f^2 \alpha \alpha_s
\left({s\over(t-m^2)}
+{t\over s}\right)                  \\
\vert T_{q\overline q}\vert ^2 &=& {128\pi^2 \over 9} Q_f^2 \alpha \alpha_s
\left({u\over(t-m^2)} + {t\over (u-m^2)}\right)
\end{eqnarray}
 where $Q_f$ and $m$ are the electric charge and the mass of a quark
$(f = u,d,s)$, and $s,\,t$ and $u$ are the standard Mandelstamm variables:
$  s=(p_1+p_2)^2,\:\: t=(p_1-k)^2,\:\: u=(p_1-k^{'})^2 $.
 
The total number of photons produced in unit volume of plasma per unit time
is given by:
\begin{eqnarray}
2\omega {dW^\gamma \over d^3 k} &=& {N_1N_2\over (2\pi)^8}
\int {d^3 k' \over2E'} \tilde n(E') \int {d^3p_1 \over2E_1} n(E_1)
\int {d^3p_2 \over 2E_2} n(E_2) \times \nonumber \\
  & & \delta^{4}(p_1+p_2-k-k') \sum_{a,b} |T_{ab}|^2 \label{Wg}
\end{eqnarray}
 where $\omega$ is the photon energy, $E_1,\,E_2,\,E'$ are the energies of
 the particles with momenta $p_1,\,p_2,\,p'$, respectively; $N_i$ is the number
 of degrees of freedom (spin, color) in the initial state of the reaction,
 $n(E)$ stand for Fermi or Bose distributions, and $\tilde n(E)=1-n(E)$.
 If to use an approximate Maxwell-Boltzmann formula $n(E)=\exp(-E/T)$,
 $\tilde n(E)=1$ instead of exact Fermi and Bose distributions,
 the multidimensional integration in (\ref{Wg}) may be performed analytically.
 The yield of photons per unit volume and unit time is then
\begin{equation}
N = {\pi\over 3}(\sum_f Q_f^2)\,\alpha\,\alpha_s(T)\,\mbox{ln}({2T \over m})T^4.
\label{Wtot} \end{equation}
 Here the running QCD coupling constant is
\begin{equation}
\alpha_s(T)={6\pi \over{(33-n_f)\,\mbox{ln}(\kappa T/\Lambda_{MS})}}.
\end{equation}
 According to lattice calculations, $\kappa\simeq 4$ and
$T_c/\Lambda_{MS}\simeq 1.8$.

Equation (\ref{Wg}) leads to the following pseudorapidity and the transverse
 momentum distributions of photons:
\begin{equation}
{d^2N \over {dk^2_\perp dy}}={4\over \pi^3}(\sum_f Q^2_f)\,\alpha\,\alpha_s(T)\,
T^2\,\mbox{ln}({4k_\perp T\over m^2})\exp(-{k_\perp\over T})\,\mbox{cosh}(y-y').
\label{spectrum} \end{equation}

 To calculate the total yield of photons from plasma we first integrate the
 hydrodynamical equations of motion and find the temperature at every time at
 every space point. Then we use eqs. (\ref{Wtot}) and (\ref{spectrum}).

 For comparison, the prediction of Bjorken model for the evolution of
 temperature with time is
\begin{equation}
 T=T_0 (\tau_0/\tau)^{nv_s^2}, \label{Bj}
\end{equation}
 where $n=1,\,2,\,3$ refer to one-, two- or three-dimensional expansion of
 plasma.

 The results of our calculations are seen in fig. 3. Fig. 3a shows the total
 rate of photon emission as a function of time. Figs. 3b,c show the
 distribution of the produced photons on the transverse momentum for
 $T_o = 258 MeV$ and $T_o = 319MeV$ respectively. Solid
 curves represent integrated spectra from the beginning till the moment when
 the pressure reaches zero level in the whole volume of the drop. Dashed
 curves correspond to the emission of photons in accord to Bjorken scenario.
Dotted curves refer to photons originating
 from decays of final-state hadrons, as is estimated in \cite{ax}.

 Variations in the model parameters may shift the presented numbers but cannot
 smash the general picture. The conclusion on the long lifetime of plasma and
 on the enhanced yield of photons remains valid within a wide range of model
 assumptions.\\~\\
\begin{center}
\begin{bf}
 6. Conclusions\\
\end{bf}
\end{center}

We have established a novel mechanism of expansion and cooling of QGP, which
consists in multiple propagation of rarefaction and compression waves trough
the plasma. Since the speed of these waves is almost constant and equal to
the speed of sound, the time needed to cool plasma increases with the size of
QGP drop.

Even if to allow the existence of states with negative pressure, the
hydrodynamical evolution of plasma cannot lead the pressure below a definite
limit, which is determined by the bag constant in the equations of state:
$p_{min}\simeq -B/2$.

Since this degree of overcooling is insufficient for supersonic phase
transitions, hadronization may only occur due to rather low processes like
deflagration or equilibrium transition.
As a consequence, the lifetime of plasma extends considerably.

Both the slow cooling mechanism and slow hadronization process result
in a longer period of photon radiation. The total yield of thermal photons
gets comparable with or even exceeds the contribution from hadron decays.

The enhanced yield of photons is a fundamental result, which may be proven
by a direct experimental measurement.
The impossibility of supersonic phase transitions from plasma to hadrons
puts some restrictions on the proportion between different hadrons and,
hopefully, may also be experimentally tested.

\newpage
Figure captions \\~\\
{\bf Fig. 1} \\
Pressure as a function of the coordinate $r$ along the radius of a QGP drop,
plotted for different times (t). The initial radius of the drop, 
the pressure and
the velocity are: $r_0 = 3~fm$, $p_0 = 6.5~GeV/fm^3$, $v_0=0$ and
$B=0.2~GeV/fm^3$.
Mind the different ordinate scale in different plots (a-l).
\\~\\~\\
{\bf Fig. 2} \\
The process of expansion and cooling of a QGP drop with the account of
deflagration.
\\~\\~\\
{\bf Fig. 3} \\
{\bf a} -- Total rate of photon emission, $T_0=258~MeV$. \\
{\bf b} -- Transverse momentum distribution of photons.\\
           Solid curve - calculations within the present model,\\
           Dashed curve - calculations according to Bjorken model,\\
           Dotted curve - background from decays of hadrons.\\
           The initial temperature $T_0=258~MeV$.\\
{\bf c} -- The same as in {\bf b}, but for $T_0=319~MeV$.
\end{large}
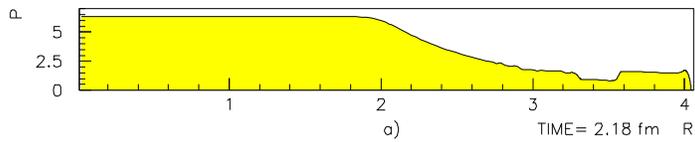
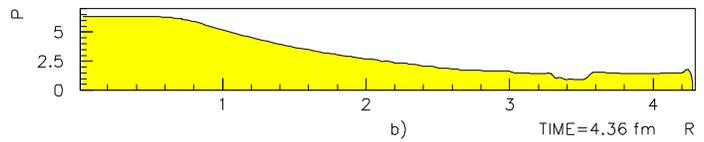
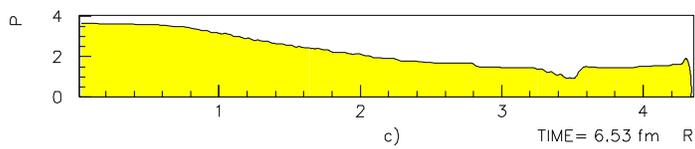
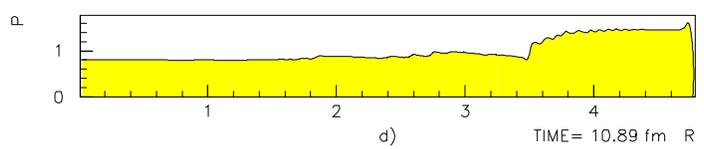
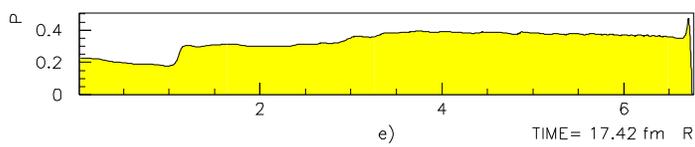
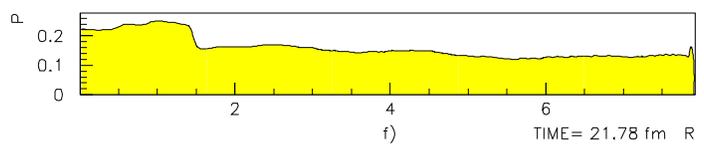
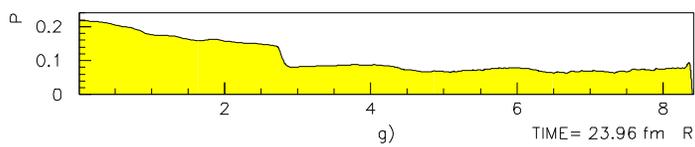
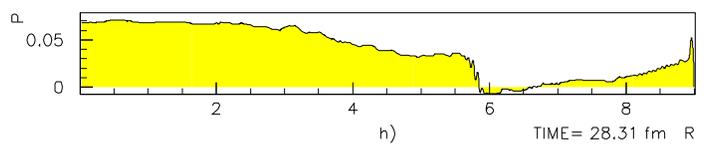
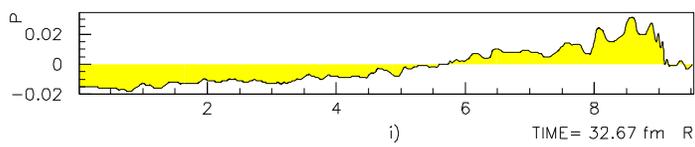
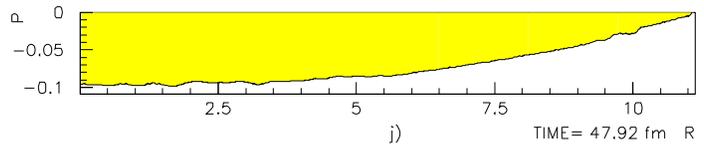
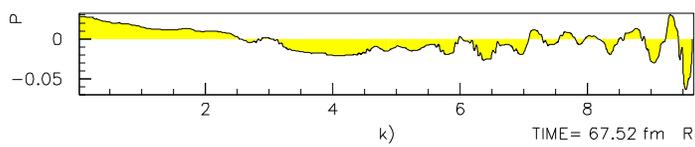
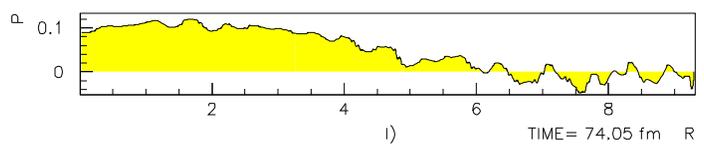
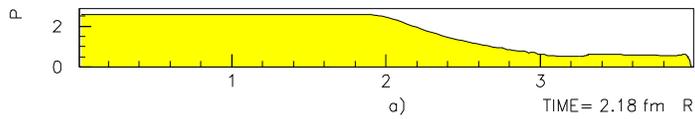
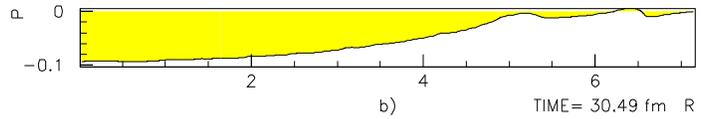
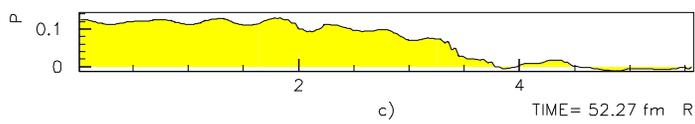
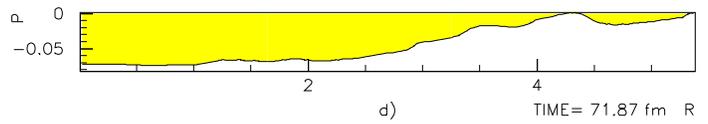
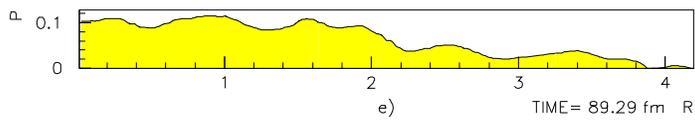
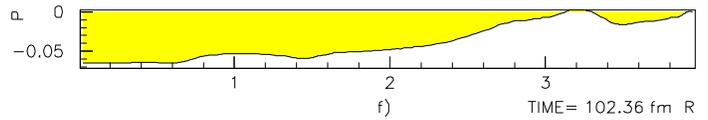
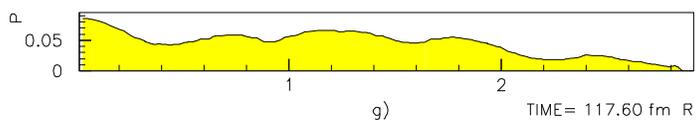
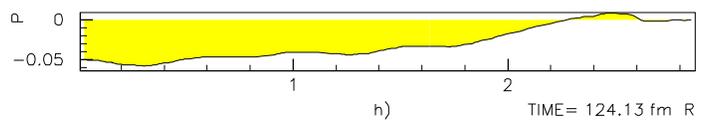
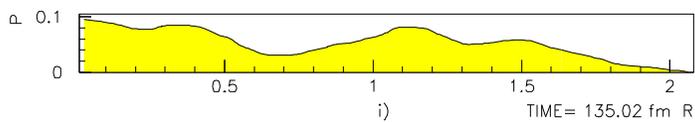
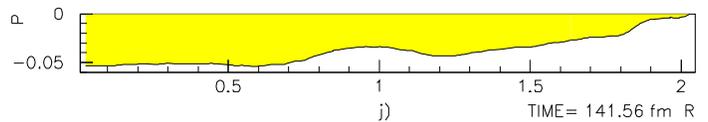
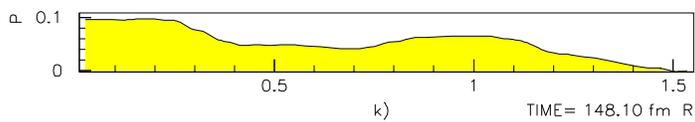
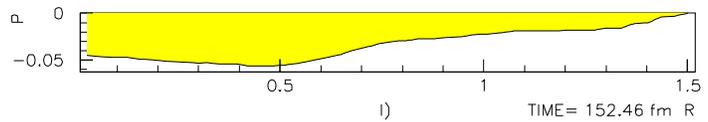
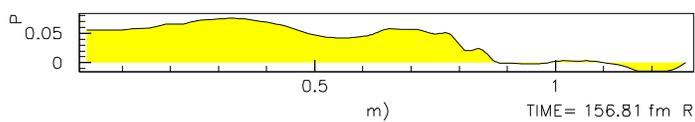
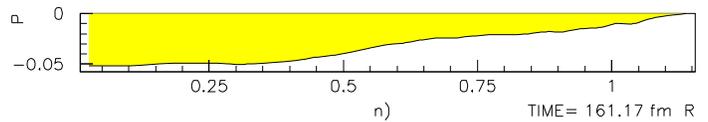
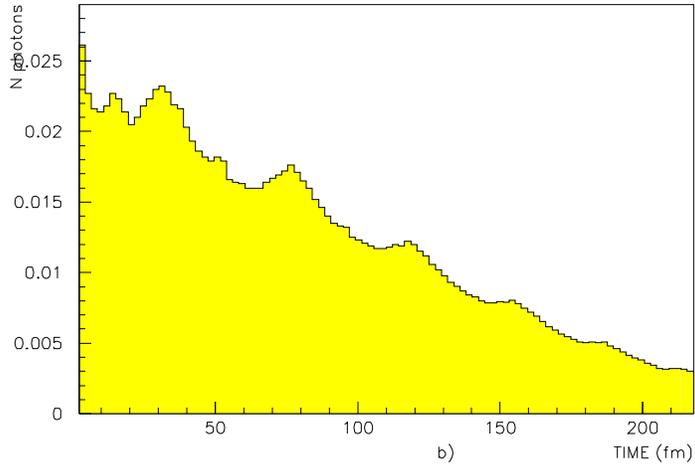
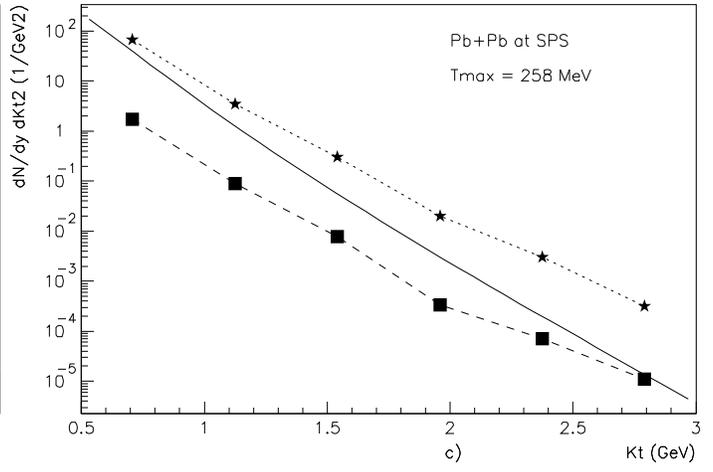
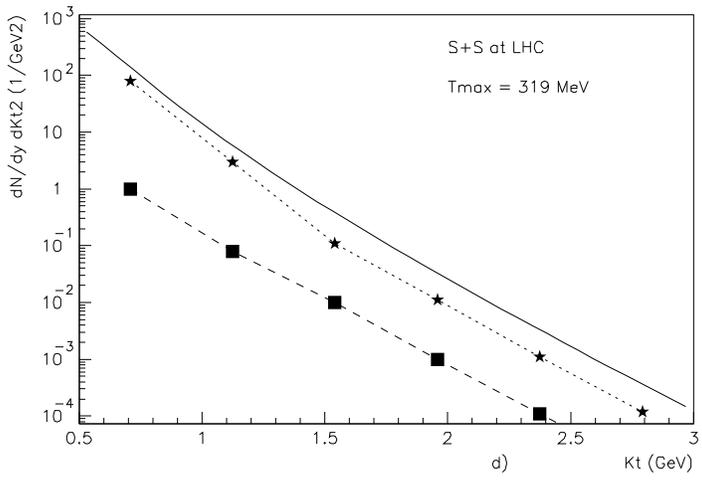
\end{document}